\documentclass[%
 reprint,
superscriptaddress,
 amsmath,amssymb,
 aps,
prapplied,
]{revtex4-1}
\usepackage{textcomp}
\usepackage{textgreek}
\usepackage{graphicx}
\usepackage{dcolumn}
\usepackage{bm}
\usepackage{lipsum}  
\newcommand{\degree}{\ensuremath{^\circ}}
\usepackage{float}
\usepackage{multirow}
\usepackage{comment}
\usepackage{natbib}
\usepackage{xcolor}

\newcommand{\picfactor}{0.28}

\newcommand{\etal}{\textit{et al.}}

\newcommand\hl[1]{%
  \bgroup
  \hskip0pt\color{black}%
  #1%
  \egroup
}

\begin{document}

\preprint{APS/123-QED}

\title{Giant magnetoelastic coupling in Love acoustic waveguide based on uniaxial multilayered TbCo$_2$/FeCo nanostructured thin film on Quartz ST-cut}

\author{Aur\'{e}lien Mazzamurro}\altaffiliation{Corresponding author: aurelien.mazzamurro@centralelille.fr}

\author{Yannick Dusch}%

\author{Philippe Pernod}%

\author{Olivier Bou Matar}

\affiliation{\mbox{Univ.~Lille,~CNRS,~Centrale~Lille,~ISEN,~Univ.~Valenciennes,~UMR~8520~-~IEMN,~LIA~LICS/LEMAC,~F-59000~Lille, France}}

\author{Ahmed Addad}
\affiliation{Univ. Lille, CNRS, INRA, ENSCL, UMR 8207 - UMET - Unit\'e  Mat\'eriaux et Transformations, F-59000 Lille, France}

\author{Abdelkrim Talbi}

\author{Nicolas Tiercelin}
\affiliation{\mbox{Univ.~Lille,~CNRS,~Centrale~Lille,~ISEN,~Univ.~Valenciennes,~UMR~8520~-~IEMN,~LIA~LICS/LEMAC,~F-59000~Lille, France}}


%


\date{\today}

\begin{abstract}
\hl{In this work, we propose a theoretical and experimental investigation of the interaction of \emph{guided} pure shear horizontal (SH) wave within a uniaxial multilayered TbCo$_2$/FeCo thin film deposited on Quartz ST-90\textdegree X cut in a delay line configuration}. We evaluate theoretically the evolution of phase velocity as a function of magnetic field and experimentally the variation of S$_{21}$ transmission coefficient (amplitude and phase). \hl{An equivalent piezomagnetic model based on pure magnetoelastic coupling was used (developed allowing us) to calculate the elastic stiffness constants of the multilayer as a function of the bias magnetic field. The model was also implemented for acoustic waves dispersion curves calculation. We show that the evolution of the phase velocity with respect to the bias magnetic field is dominated by the $C_{66}$ elastic stiffness constant as expected for the case of shear horizontal surface acoustic wave.} In the fabricated device, both fundamental and third harmonic shear mode are excited at 410~MHz and 1.2~GHz, respectively. For both modes, the theoretical and experimental results are in agreement. \hl{At 1.2~GHz the guiding of the acoustic wave in the ferromagnetic thin film enhances the sensitivity to the bias magnetic field with a maximum phase velocity shift close to 2.5~\% and an attenuation reaching 500~dB/cm, for a sensitivity as high as 250~ppm/Oe, which is better than what has been reported in literature so far. We also report that, from a specific ratio between the thin film thickness and the acoustic wavelength, the bias magnetic field can induce a breaking of the acoustic wave polarization, leading to an acoustic mode conversion.}
\end{abstract}

\maketitle



\section*{Introduction}
Magnetic field sensing is widely used for numerous applications ranging from navigation and positioning systems such as inertial units, magnetic anomaly detection, to electrical current sensors or biomagnetic signals sensors such as magneto-cardiography, magneto-encephalography or hepatic iron concentration. In such applications, common demands concerning the sensors capabilities are: room temperature operation, compact size for high spatial resolution or limited available space, and low power consumption. \hl{For the past few years, various approaches have emerged to fill the aforementioned criteria. Thin film/plate based magnetoelectric resonators \cite{Zhai2007, Ou-Yang2019, Nan2013, Hui2015, Piorra2013} show encouraging results but their limited frequency range due to down-scaling constraints reduces their potential. In the meantime, Surface Acoustic Wave (SAW) based devices have emerged as a promising technology in magnetic field sensing with the use of magnetostrictive materials being able to convert magnetic energy to mechanical energy and vice-versa.} When biased with a magnetic field, strain/stress is induced in the magnetostrictive thin film due to the natural alignment of the magnetization along the external perturbation leading to elastic strain. The modification of the elastic properties of magnetostrictive composites with respect to a bias magnetic field is known as \textDelta E/G-effect \cite{Sarkozi2000, Ludwig2002, robbins_simple_1988} \hl{which reflects bulk or shear modulus variation, respectively. Typically, SAW devices are operated in a delay line configuration \cite{Wang2018, Kittmann2018, Li2012, Forester1978, Webb1977, elhosni_theoretical_2014, zhou_multilayer_2014} or in a resonator structure with one or two ports \cite{Polewczyk2017, Smole2004, Liu2019, Kadota2012}. SAW devices are preferred for RF applications since down-scaling and thus, frequency increase, is much easier compared to magnetoelectric based resonators. In the former case, coupling between dynamic strain and magnetization was investigated by various research teams through either resonant or non-resonant interaction. In the latter, the magnetoelastic coupling induces the shift of the acoustic wave dispersion curves, whereas in resonant coupling, dynamic strain is used to induce the magnetization precession corresponding to the crossing between magnon and phonon dispersion curves. In previous studies reported in literature, non-resonant coupling was investigated with various ferromagnetic materials such as FeGa, Ni, TbFe$_2$, TbCo$_2$, FeCo or FeCoSiB using leaky or surface acoustic waves \cite{elhosni_experimental_2015, Li2012, zhou_multilayer_2014,dreher_surface_2012} showing sagittal (including Rayleigh waves) or transverse polarization with a penetration depth reaching few to several wavelengths, reducing accordingly the sensitivity.} In particular, Li \etal{} \cite{Li2012} reported the design of a magnetic field sensor based on 500 nm thick FeGa film deposited on Quartz ST-cut showing a maximum velocity shift of 0.64\% at 158 MHz. Zhou \etal{} \cite{zhou_multilayer_2014} conducted the experimental and theoretical study of a SAW delay line based on multilayered TbCo$_2$/FeCo nanostructured thin film deposited on Y-cut LiNbO$_3$. A very good agreement was reached between theoretical predictions and experimental data, both for Rayleigh and shear mode with a maximum acoustic wave velocity shift close to 0.2\%. Elhosni \etal{} \cite{elhosni_experimental_2015} investigated experimentally and theoretically the use of Ni and CoFeB as sensing materials in a delay line configuration on LiNbO$_3$ YX-128\textdegree{} coated with ZnO. A maximum sensitivity of 16 ppm/mT was reached with CoFeB at 460 MHz. Kittmann \etal{} \cite{Kittmann2018} presented a FeCoSiB thin film coated Love type SAW device for magnetic field sensing with a very low magnetic noise level of 100 pT/$\sqrt{Hz}$. SAW resonator structures were also investigated and noticeable results were obtained: Smole \etal{} \cite{Smole2004} proposed a one port resonator in a FeCoSiB/ZnO structure with a tuning range of about 1.2\% at 1.2 GHz. Kadota \etal{} \cite{Kadota2012} showed the first the excitation of a pure shear horizontal surface acoustic wave on Quartz ST-X90\textdegree{} cut to design resonators as magnetic field sensor. A one port resonator composed of Ni IDTs with a sensitivity to magnetic field close to 8 ppm/Oe was designed. More recently, Liu \etal{} \cite{Liu2019} reported classic Love acoustic waveguide in a one port resonator with SiO$_2$ layer and FeCoSiB as magnetostrictive layer with a relative frequency shift of 0.5\% at 220 MHz. \\
The coupling between dynamic strain and magnetization is also interpreted as ferromagnetic resonance (FMR) by some research teams \cite{Labanowski2017, Duquesne2019, thevenard_surface-acoustic-wave-driven_2014, dreher_surface_2012, weiler_elastically_2011}. Thevenard \etal{} \cite{thevenard_surface-acoustic-wave-driven_2014} showed the dynamic strain effect on a magnetic semiconductor coated with a ZnO layer and obtained a velocity shift close to 0.007\% and an attenuation close to 1 dB/cm at 549 MHz. Dreher \etal{} \cite{dreher_surface_2012} investigated acoustic wave driven FMR (ADFMR) in a Ni thin film using a SAW delay line configuration on LiNbO$_3$ at high frequency. At 170 MHz, an attenuation close to 1 dB/cm was obtained, reaching 160 dB/cm at 2.24 GHz. Labanowski \etal{} also investigated ADFMR in a Ni thin film in a delay line configuration on Y-cut LiNbO$_3$ and an attenuation of 500 dB/cm at 2.5 GHz for a 20 nm thick thin film was claimed by the authors. In these works, although a resonant coupling between magnetization and dynamic strain is claimed, the relative acoustic wave velocity shifts observed with respect to the bias magnetic field remain weak compared to non resonant coupling.

\hl{In summary, most of the studies concerning the coupling between dynamic strain and magnetization focus on leaky-type or surface acoustic wave to induce magnetization precession or pure magnetoelastic effect. The study of the true nature of this coupling presents a major challenge owing to the complex acoustic wave polarization generally used and the magnetization state of the ferromagnetic thin film. The studies concerning resonant coupling show a great interest in magnon-phonon interaction at a fundamental level, but the sensitivity with respect to the magnetic field remains weak compared to pure magnetoelastic coupling based devices for magnetic field sensor applications. However, the latter devices have also a limited sensitivity due to the evanescent nature of the employed acoustic waves, a non negligible part being radiated into the substrate. With Love wave devices, the confinement of the acoustic wave is better but only a small part travels in the ferromagnetic layer coated on the silicon dioxide. In order to enhance the sensitivity, improving the confinement of the acoustic wave into the ferromagnetic thin film is thus required. 
In this work, we therefore propose to study the interaction of \emph{guided} pure shear horizontal (SH) wave within uniaxial multilayered TbCo$_2$/FeCo thin film (in-plane magnetization) deposited on Quartz ST-90\textdegree X cut in a delay line configuration. In our case, the Love wave condition is fulfilled thanks to the chosen ferromagnetic material showing a very high density and softness. Besides enhanced sensitivity, the purpose of this work is to fully understand the SAW coupling occurring in such ferromagnetic thin film which is rendered easier thanks to the pure shear horizontal nature of the chosen acoustic wave.}

The paper is structured as follows: the magnetoelastic coupling between pure SH wave and a uniaxial nanostructured multilayered magnetostrictive thin film is first described through the equivalent piezomagnetic model in Section I. Measurements performed on the fabricated SAW devices are then reported in section II, as well as the comparison with the theoretical data to validate the piezomagnetic model.

\section{Theoretical considerations}
\subsection{SH surface wave in Quartz ST-90\textdegree{} X cut}
The Quartz ST-cut is a popular piezoelectric material used for decades to design SAW filters, sensors or temperature stabilized SAW devices. SAW devices are characterized by three main parameters: phase velocity, electromechanical coupling coefficient ($\textrm{K}^\textrm{2}$), and temperature coefficient of frequency. These parameters usually depend on the the substrate orientation and wave propagation direction\hl{which is here referenced by the angle $\Psi$, arbitrarily chosen relatively to the x-axis.} A numerical method that was developed in a previous work \cite{noauthor_legendre_nodate} to compute the dispersion curves and mode shapes of elastic waves in layered piezoelectric-piezomagnetic composites, is used to obtain the phase velocity and the mode shapes existing in Quartz ST-cut (Euler angles (0\textdegree{}, 132.75\textdegree{}, $\Psi$)) depending on the propagation direction $\Psi$.

\begin{figure}[]
    \centering
    \includegraphics[width=\columnwidth]{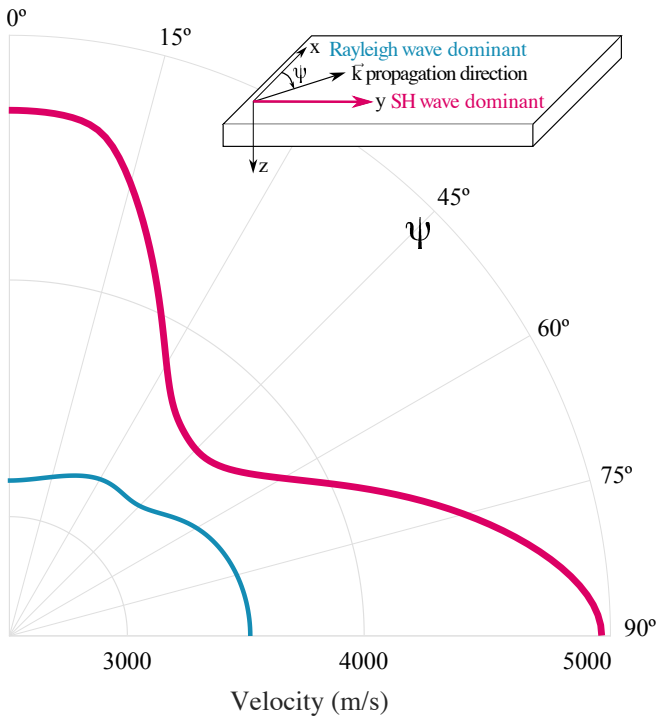}
    \caption{\hl{Phase velocity of the Rayleigh wave (blue) and the pure SH mode (magenta) as a function of the propagation direction $\Psi$.}}
    \label{fig:modes}
\end{figure}

\hl{The phase velocities of the pure SH mode (magenta) and the Rayleigh wave (blue) are reported in FIG.~\ref{fig:modes} as a function of the direction of propagation $\Psi$.}

\begin{figure}[]
    \centering
    \includegraphics[width=\columnwidth]{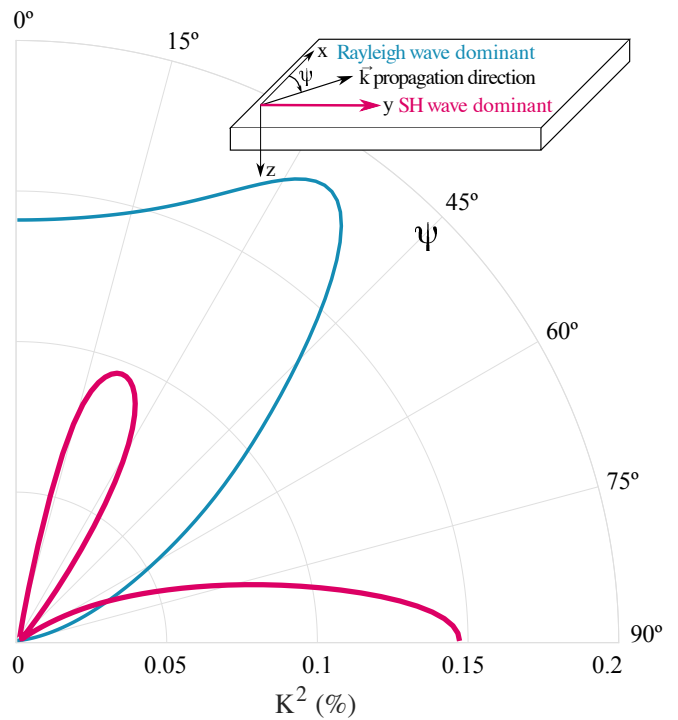}
    \caption{\hl{Electromechanical coupling coefficient $K^2$ of the Rayleigh wave (blue) and the pure SH mode (magenta) as a function of the propagation direction $\Psi$.}}
    \label{fig:k2}
\end{figure}

\begin{figure}[]
    \centering
    \includegraphics[width=\columnwidth]{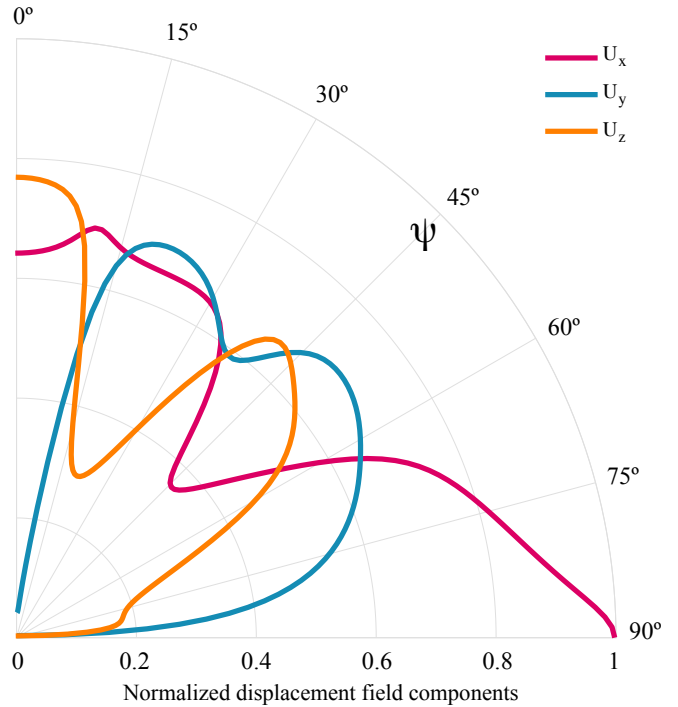}
    \caption{Displacement field of the SH mode as a function of the propagation direction $\Psi$.}
    \label{fig:disp_field}
\end{figure}

\begin{figure}[]
    \centering
    \includegraphics[width=\columnwidth]{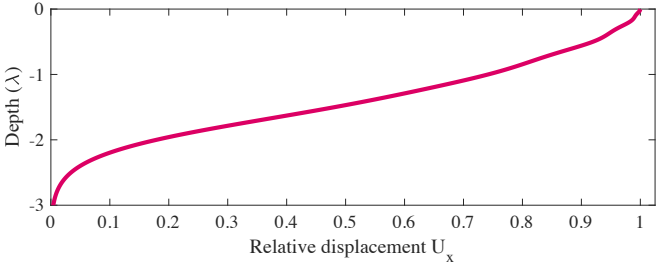}
    \caption{Relative displacement amplitude $U_x$ and normalized penetration depth for the SH mode ($\Psi = 90\degree$).}
    \label{fig:Ux}
\end{figure}

In order to assess the predominant mode of propagation in such material depending on the in-plane direction $\Psi$, the electromechanical coupling coefficient $\textrm{K}^\textrm{2}$ is then computed for both modes and is given by 
\begin{equation}
    K^2 = 2 \frac{v_{co}-v_{cc}}{v_{co}},
\end{equation}
where $v_{co}$ and $v_{cc}$  are the phase velocities of the open and short circuit boundary condition. $K^2(\Psi)$ is shown in FIG.~\ref{fig:k2} for the Rayleigh wave (blue) and the pure SH mode (magenta). The Rayleigh wave is propagating at 3158 m/s for $\Psi~=~0\degree$ with $\textrm{K}^\textrm{2}$~=~0.14\%. For $\Psi~=~90\degree$, the solely coupled mode is the pure SH mode (magenta) and shows a phase velocity of 5000~m/s and a $\textrm{K}^\textrm{2}$~=~0.15\%. 
The relative displacement field ($U_x$, $U_y$, $U_z$) of this mode is computed along the propagation direction $\Psi$ and reported in FIG.~\ref{fig:disp_field}. It is evident that this mode turns into a pure shear horizontal mode when $\Psi~=~90\degree$, the shear horizontal component $U_x$ being the only one to exist. FIG.~\ref{fig:Ux} shows that $U_x$ decreases almost to zero within a depth of 3$\lambda$. Thus, the SH mode is intrinsically a guided surface acoustic wave, and most of the energy is trapped on the propagation surface with no spurious coupling. 

Therefore, the SH mode is only carrying an in-plane unidirectional shear displacement, perpendicular to the direction of propagation. In the following section, the interaction with a magnetostrictive material (multilayered nanostructured TbCo$_2$/FeCo) is investigated through the equivalent piezomagnetic model. 

\subsection{Equivalent piezomagnetic model}
 \label{eqpiezomag}
\begin{figure}
     \centering
     \includegraphics{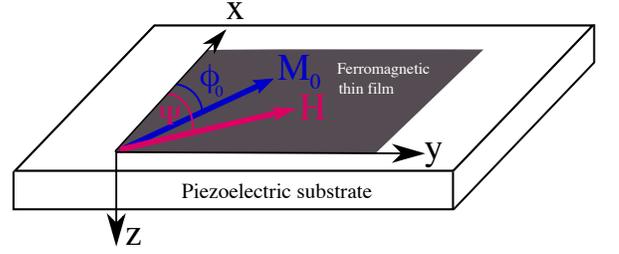}
     \caption{System coordinates used in the piezomagnetic model}
     \label{fig:mag_ferro}
 \end{figure}

 The equivalent piezomagnetic model was developed previously by the authors to assess magnetoelastic coupling in piezo-electro-magnetic composites (for further details, see \cite{noauthor_legendre_nodate, zhou_multilayer_2014}). The piezomagnetic equations are obtained by considering a magnetoelastic wave in a ferromagnetic thin film deposited on a piezoelectric substrate or membrane (see Fig.~\ref{fig:mag_ferro}), and linearizing the coupled equations for the mechanical and magnetic systems (Newton's equation of motion and Landau-Lifshitz equation) around a ground state position of the magnetization (depending on the direction and magnitude of the bias magnetic field). In the coordinate system described in FIG.~\ref{fig:mag_ferro} the piezomagnetic equations are given by
\begin{align}
\rho \frac{\partial^2 u_i}{\partial t^2} &= \frac{\partial \sigma_{ij}}{\partial x_j}, \\
\frac{\partial b_i}{\partial x_i} &= \frac{\partial (\mu_0(h_i+m_i))}{\partial x_i},
\end{align}
where $\rho$ is the density of the ferromagnetic thin film, $u_i$ is the $i^\textrm{th}$ component of the particle displacement, $x_i$ denotes the Eulerian coordinates (Einstein's summation convention is used, \emph{i,j,k}, and \emph{l} = 1,2,3 or equivalently $x,y,z$) and
\begin{align}
    \label{eq:sigma}
    \sigma_{ij} &= (C_{ijkl}  + \Delta C_{ijkl})\frac{\partial u_k}{\partial x_l} - q_{lij}h_l, \\
    \label{eq:b}
    b_i &= q_{ikl}\frac{\partial u_k}{\partial x_l}  + \mu_{il}h_l,
\end{align}
with $C_{ijkl}$ the elastic stiffness constants and where the effective magnetic permeability $\mu_{ij}$ and elastic stiffness constants corrections $\Delta C_{ijkl}$ are given by 
\begin{align}
    \mu_{il} &= \mu_0(\delta_{il}+\chi_{il}),\\
    \Delta C_{ijkl} &= b_{ijmn}(M^{0}_n q_{mkl}+ M^{0}_m q_{nkl})
    \label{eq:delta_cijkl}
\end{align}
with $b_{ijkl}$, the magnetoelastic constants ($b_{1111} = b^{\gamma,2}$ when using Callen's notations \cite{EARLR.CALLEN1963}). The expressions of the piezomagnetic constants $q_{ijk}$ and the magnetic susceptibility $\chi_{il}$ can be found in \cite{bou_matar_band_2012}.
 
\begin{figure}
    \centering
    \includegraphics[width=\columnwidth]{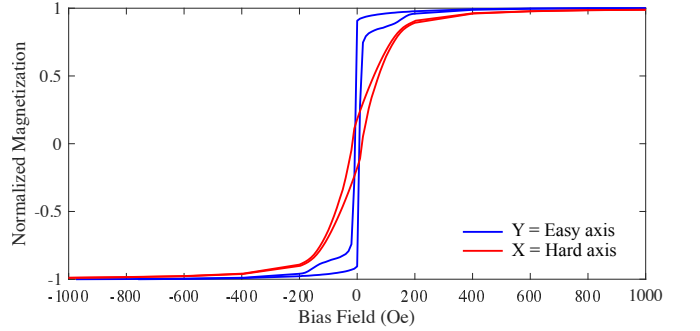}
    \caption{TbCo$_2$/FeCo thin film magnetization characteristics measured along the easy axis (blue) and the hard axis (red).}
    \label{fig:mag}
\end{figure}

In the following, we consider a uniaxial multilayered 25~x~[TbCo$_{2(3.7nm)}$/FeCo$_{(4nm)}$] nanostructured thin film sputtered under a bias magnetic field ($C_{11}$ = 110.9 GPa, $C_{44}$ = 31.3 GPa, $\rho$ = 9250 kg.$\textrm{m}^{\textrm{-3}}$, E = 80 GPa, $b^{\gamma,2}$ = -4.5 MPa, $\textrm{H}_\textrm{A}$ = 200 Oe). \hl{The in-plane magnetization was characterized using a Vibrating Sample Magnetometer (VSM), when applying the magnetic field along the x and y-axis. A clear uni-axial anisotropy is observed as shown on FIG.~\ref{fig:mag}. The x and y-axis are respectively the Hard (red) and Easy (blue) magnetic axis. The anisotropy field $\textrm{H}_\textrm{A}$ is estimated to be around 200~Oe. In this study, the bias magnetic field is applied along the hard axis, while the acoustic wave propagates along the y-axis.} The magnetoelastic coupling induces anisotropy in the magnetostrictive thin film and thus, it cannot be considered as isotropic. The magnetization is supposed to remain always in the plane of the film. For calculations, the magnetic field is applied perpendicular to the easy axis of the magnetostrictive thin film. Its effective elastic constants $C_{ijkl}$, piezomagnetic constants $q_{ijk}$ and relative magnetic permeability constants $\mu_{ik}$ are computed and reported in Appendix~\ref{app:cij}, FIG.~\ref{fig:qij} and FIG~\ref{fig:muij}, respectively. 

The expressions of the elastic stiffness corrections $\Delta C_{ijkl}$ to $C_{ijkl}$ due to $\Delta$E-effect, are given in Table~\ref{tab:deltac}. In these expressions, $b^{\gamma,2}$ is the magnetoelastic coupling coefficient of the isotropic thin film, and 
\begin{align}
    U'_{\theta \theta} &= \mu_0 M_s (H_A (1-sin^2\phi_0) + H cos(\phi_0-\psi)+H_{me}), \nonumber\\
    U'_{\phi \phi} &= \mu_0 M_s (H_A(1-2sin^2\phi_0)+H cos(\phi_0-\psi)+H_{me}),
\end{align}
where $H_{me}=(b^{\gamma,2})^2/(\mu_0 M_s C_{44})$ is the magnetoelastic field, $\phi_0$ and $\psi$ the angle respect to the x-axis with magnetization and magnetic field, respectively. Therefore, the elastic stiffness tensor of the magnetostrictive thin film is modified when biased with a magnetic field and the elastic stiffness constants are modified as follows
\begin{table}
\caption{Non zero components of the effective elastic stiffness correction tensor.}
\begin{ruledtabular}
    \begin{tabular}{@{}ll@{}}
        $\Delta C_{11}$ = $\frac{-4(b^{\gamma,2})^2 \textrm{cos}^2\phi_0 \textrm{sin}^2\phi_0}{U'_{\phi \phi}}$ &
         $\Delta C_{16}$ = $\frac{(b^{\gamma,2})^2 \textrm{sin}(4\phi_0)}{2U'_{\phi \phi}}$\\[10pt]
         $\Delta C_{44}$ = $\frac{-(b^{\gamma,2})^2 \textrm{sin}^2\phi_0}{U'_{\theta \theta}}$ &
         $\Delta C_{45}$ = $\frac{-(b^{\gamma,2})^2 \textrm{sin}\phi_0 \textrm{cos}\phi_0}{U'_{\theta \theta}}$\\[10pt]
         $\Delta C_{55}$ = $\frac{-(b^{\gamma,2})^2 \textrm{cos}^2\phi_0}{U'_{\theta \theta}}$ &
         $\Delta C_{66}$ = $\frac{-(b^{\gamma,2})^2 \textrm{cos}^2 (2\phi_0)}{U'_{\phi \phi}}$\\
    \end{tabular}
    \label{tab:deltac}
\end{ruledtabular}
\end{table}

\begin{equation}
\begin{bmatrix}
\Delta C_{11} & -\Delta C_{11} & 0 & 0 & 0 & \Delta C_{16}\\
-\Delta C_{11} & \Delta C_{11} & 0 & 0 & 0 & -\Delta C_{16}\\
0 & 0 & 0 & 0 & 0 & 0\\
0 & 0 & 0 & \Delta C_{44} & \Delta C_{45} & 0\\
0 & 0 & 0 & \Delta C_{45} & \Delta C_{55} & 0\\
\Delta C_{16} & -\Delta C_{16} & 0 & 0 & 0 & \Delta C_{66}
\end{bmatrix}.
\label{eq:deltac}
\end{equation}
The dependency of the overall elastic stiffness constants $C_{ijkl}$ with respect to the magnetic field is reported in Appendix~\ref{app:cij}.

\begin{figure*}
\includegraphics[width=\textwidth]{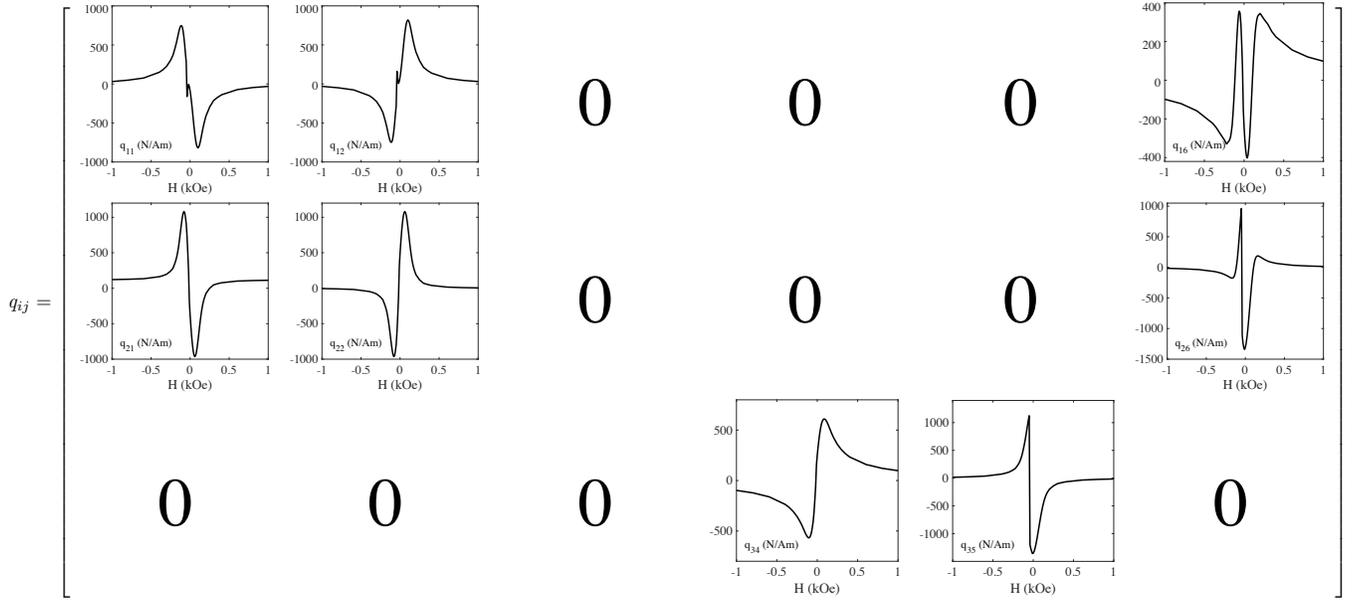}
\caption{Magnetic field dependency of the piezomagnetic constant tensor when a bias magnetic field is applied along the hard axis of a multilayered TbCo$_2$/FeCo thin film on Quartz ST-X90\textdegree{} cut.}
\label{fig:qij}
\end{figure*}

\begin{figure}
\includegraphics[width=\columnwidth]{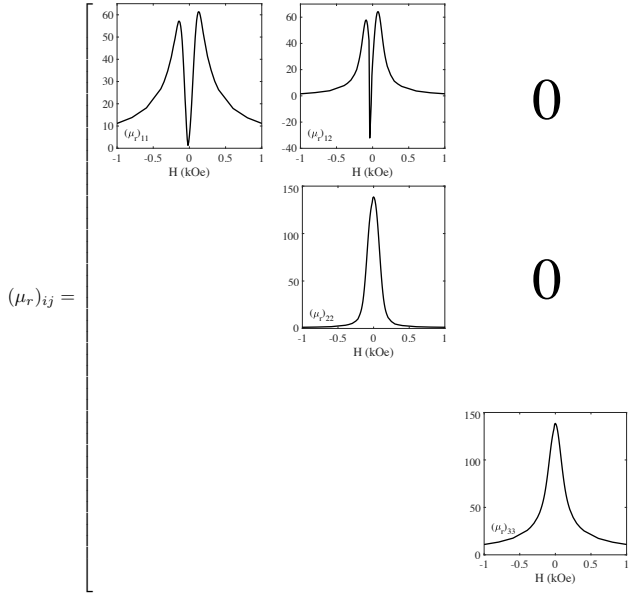}
\caption{Magnetic field dependency of the relative magnetic permeability constant tensor when a bias magnetic field is applied along the hard axis of a multilayered TbCo$_2$/FeCo thin film on Quartz ST-X90\textdegree{} cut.}
\label{fig:muij}
\end{figure}

\subsection{Analytical resolution for Shear Horizontal wave in anisotropic media}
In order to enhance the sensitivity of the surface acoustic wave to the magnetostrictive thin film strain when biased with a magnetic field, there are essentially three approaches: First, using bulk waves to exploit $C_{11}$ or $C_{22}$ variations, through coating of thin piezoelectric plates, or interaction with a Rayleigh wave, mainly sensitive to $C_{11}$. Second, manufacturing magnetostrictive thin films with out-of-plane anisotropy to use $C_{33}$ variations, which is not considered here, since the magnetization remains in-plane \hl{because of strong demagnetizing effects}. Third, using shear surface acoustic waves to take advantage of the $C_{66}$ dependency with the magnetic field. The latter option was retained, and Quartz ST-X90\textdegree{} cut was chosen for its pure shear horizontal wave to exploit the induced shear stress/strain of the magnetostrictive thin film. 

\begin{figure}[]
    \centering
    \includegraphics[width=\columnwidth]{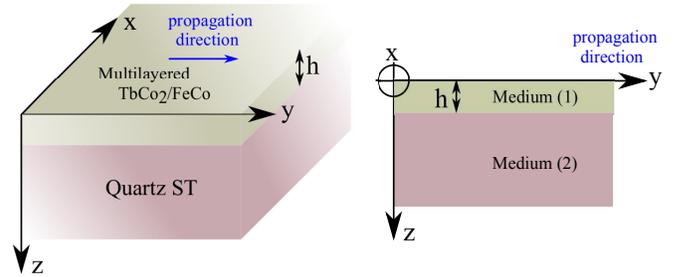}
    \caption{System coordinates used for the Love wave device.}
    \label{fig:lw_coords}
\end{figure}

The velocity dependency with a bias magnetic field of the pure SH mode propagating in this piezo-electro-magnetic composite can be easily obtained analytically for anisotropic media considering that both the elastic stiffness matrix of the substrate and ferromagnetic layer are included in the monoclinic system case. The coordinate system and geometry used in this study are depicted in FIG~\ref{fig:lw_coords}. For simplicity, the piezoelectricity is neglected. The magnetostrictive layer ($0 < z < h$) is designated as medium (1) with displacement ${u}_{1}^{(1)}(y, z, t)$, density $\rho_1$ and elastic constants ${c}_{ij}$ and the half-space (z $\geq$ h) is designated as medium (2) with displacement ${u}_{1}^{(2)}(y, z, t)$, density $\rho_2$ and elastic constants ${d}_{ij}$. Considering a Love wave ($u_2 = u_3 = 0$) propagating in the positive y-direction with a phase velocity $c$, and assuming a $e^{-j(\omega t - ky)}$ dependence, the displacement field $u_1$ can be expressed as
\begin{equation}
    u_1 = f(z)e^{jk(y - ct)}.
    \label{eq:u1}
\end{equation}

The Christoffel's tensor ruling wave propagation in solid media is therefore reduced to 

\begin{equation}
    (\Gamma_{11}-\rho c^2)u_1 = 0
    \label{eq:christoffel}
\end{equation}
with $\rho$ the material density, $c=\omega/k$ the phase velocity and $\Gamma_{il} = c_{ijkl}n_jn_k$, $n_{m}$ being the components of the unit vector in the wave propagation direction.

The overall equation ruling wave propagation in both materials (condensed notation) is therefore given by
\begin{equation}
 \rho_{1,2} \frac{\partial^2 u_1}{\partial t^2} = (c,d)_{66}\frac{\partial^2 u_1}{\partial y^2} + 2(c,d)_{56}\frac{\partial^2 u_1}{\partial y \partial z} + (c,d)_{55}\frac{\partial^2 u_1}{\partial z^2}
 \label{eq:gen_eq}
\end{equation}

The general solution of Eq.~\ref{eq:gen_eq} for the medium (1) is
\begin{equation}
    {u}_{1}^{(1)} = (A_1e^{jkb_1z}+B_1e^{-jkb_2z})e^{jk(y-ct)},
    \label{eq:sol_u1}
\end{equation}
where $A_1$, $B_1$ are arbitrary constants and 
\begin{align}
    b_1 &= (\sqrt{A} - {c}_{56})/{c}_{55}, \quad b_2 = (\sqrt{A} + {c}_{56})/{c}_{55} \nonumber \\
    A &= {c}_{55}(\rho_1 c^2 - {c}_{66}) + {c}_{56}^{2}
    \label{eq:A}
\end{align}

Likewise, given $\lim_{z\to\infty} {u}_{1}^{(2)} = 0$, the displacement field ${u}_{1}^{(2)}$ can be expressed as
\begin{equation}
    {u}_{1}^{(2)} = A_2e^{jkb_3z}e^{jk(y-ct)}
\end{equation}
with 
\begin{equation}
    b_3 = \frac{j\sqrt{B}-{d}_{56}}{{d}_{55}},\quad B = {d}_{55}({d}_{56}-\rho_2c^2)-{d}_{56}^{2}
    \label{eq:B}
\end{equation}
with $\Im({b_3}) > 0$ (i.e. $B>0$). The boundary conditions are
\begin{align}
    {\tau}_{13}^{(1)} &= 0 & \text{at z = 0 (free surface)},\\
    {u}_{1}^{(1)} &= {u}_{1}^{(2)} & \text{at z = H (displacement continuity)},\\
    {\tau}_{13}^{(1)} &= {\tau}_{13}^{(2)} & \text{at z = H (stress continuity)}.
    \label{eq:cl}
\end{align}

But
\begin{align}
    {\tau}_{13}^{(1)} &= {c}_{56}\frac{\partial {u}_{1}^{(1)}}{\partial y} + {c}_{55}\frac{\partial {u}_{1}^{(1)}}{\partial z}\\
    {\tau}_{13}^{(2)} &= {d}_{56}\frac{\partial {u}_{1}^{(2)}}{\partial y} + {d}_{55}\frac{\partial {u}_{1}^{(2)}}{\partial z}
\end{align}

(\ref{eq:sol_u1}) to (\ref{eq:cl}) yield
\begin{align}
    A_1 - B_1 &= 0, \nonumber \\
    A_1e^{jkb_1 h} + B_1e^{-jkb_2 h} &= A_2e^{jkb_3 h} \nonumber \\
    A_1e^{jkb_1 h} - B_1e^{-jkb_2 h} &= j\sqrt{\frac{B}{A}}A_2e^{jkb_3 h}
    \label{eq:cl1}
\end{align}

Eliminating $A_1$, $B_1$, and $A_2$ in (\ref{eq:cl1}) gives

\begin{equation}
    \tan{\theta} = \sqrt{\frac{B}{A}}
\end{equation}
with 
\begin{equation}
    \theta = \frac{(b_1+b_2)k h}{2} = \frac{\sqrt{A}}{{d}_{55}}k h   
\end{equation}

It can be shown that for Love waves to exist, $A>0$, $B>0$. Therefore, using (\ref{eq:A}) and (\ref{eq:B}) the Love wave dispersion relation may be written as
\begin{equation}
\boxed{
    \tan{(\gamma\chi_{1} k h)} = \sqrt{\frac{{d}_{55}{d}_{66}}{{c}_{55}{c}_{66}}}\frac{\chi_{2}}{\chi_{1}} 
    }
    \label{eq:lovewave}
\end{equation}
with 
\begin{align}
    {\gamma}^{2} = \frac{{c}_{66}}{{c}_{55}}, \quad \epsilon_{1} = \frac{{c}_{56}^{2}}{{c}_{55}{c}_{66}}, \quad \epsilon_{2} = \frac{{d}_{56}^{2}}{{d}_{55}{d}_{66}} \nonumber \\
\end{align}
and the variables $\chi_1$ and $\chi_2$ are given by
\begin{align}
    \chi_{1} = \sqrt{\frac{c^2}{{V}_{1}^{2}}-1+\epsilon_{1}}, \quad \chi_{2} = \sqrt{1-\epsilon_{2}-\frac{c^2}{{V}_{2}^{2}}}
\end{align}

where $V_1$ and $V_2$ are the bulk shear velocity in both media and are given by
\begin{align}
    V_{1} = \sqrt{\frac{{c}_{66}}{\rho_1}}, \quad V_{2} = \sqrt{\frac{{d}_{66}}{\rho_2}} \nonumber\\
    \label{eq:vsh}
\end{align}

The conditions $A>0$ and $B>0$ imply
\begin{equation}
    \sqrt{1-\epsilon_{1}} V_{1} < c < \sqrt{1-\epsilon_{2}} V_{2}
    \label{eq:lwcond}
\end{equation}

In this paper, for the considered magnetostrictive thin film, $\epsilon_{1} = 0$, since ${c}_{56} = 0$. Moreover, it can be noticed (see Appendix~\ref{app:cij}) that it behaves as an \hl{elastically} isotropic material when saturated with a bias magnetic field, the magnetization being stuck in a well defined state, the elasticity remains unchanged. In order to take the piezomagnetic effects into account in the analytical Love wave dispersion relation, the elastic stiffness constants are rewritten using (\ref{eq:sigma}) and (\ref{eq:b}) as 
\begin{equation}
    C'_{ijkl} = {C}_{ijkl}^{H} + \frac{q_{pij} n_p n_q q_{qkl}}{\mu_{mn} n_m n_n}
\end{equation}
where $n_i$ are the components of the wave vector.
These enhanced elastic stiffness constants can be used in the Love wave dispersion relation to obtain the wave velocity dependency with respect to a bias magnetic field. \hl{Its evolution will be discussed later, along the characterization of the experimental device in Section \ref{disc}.}

\section{Experimental considerations and discussion}
\label{sec2}
\subsection{Fabricated SAW delay line}
The interaction of the multilayered magnetostrictive thin film with pure shear surface acoustic wave was also investigated experimentally. A SAW delay line was retained as acoustic waveguide. As shown in FIG~\ref{fig:dl} and FIG~\ref{fig:dispo} (left), a 300~$\mu m$ wide and 200~nm thick (0.016$\lambda$) 25~x~[TbCo$_{2(3.7\textrm{nm})}$/FeCo$_{(4\textrm{nm})}$] nanostructured uniaxial thin film was deposited by RF-sputtering onto a 500~$\mu m$ thick Quartz ST-cut substrate between two single-electrode aluminum interdigital transducers (IDTs) with a periodicity of 12~$\mu m$. The ferromagnetic thin film was deposited under a bias magnetic field to induce the preferred orientation for the magnetization, as seen on FIG.~\ref{fig:mag}. As stated before, the propagation direction of the acoustic wave is along the Easy axis (EA) of the ferromagnetic thin film.
\hl{Transmission Electron Microscopy on the multilayered nanostructured thin film has been made on a TEM FEI Themis probe corrected, and equipped with a SUPER-X-EDX detector. Imaging was done on STEM mode with the High-Angle Annular Dark-Field (HAADF) detector, with which atomic resolution imaging can be achieved (0.07nm). The resulting picture is reported in FIG~\ref{fig:dispo} (right), where each TbCo$_2$/FeCo bilayer can be clearly seen.

\begin{figure}[]
    \centering
    \includegraphics[width=\columnwidth]{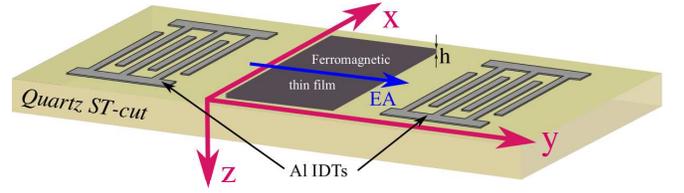}
    \caption{3D model of the fabricated Love wave device.}
    \label{fig:dl}
\end{figure}

\begin{figure}[]
    \centering
    \includegraphics[width=\columnwidth]{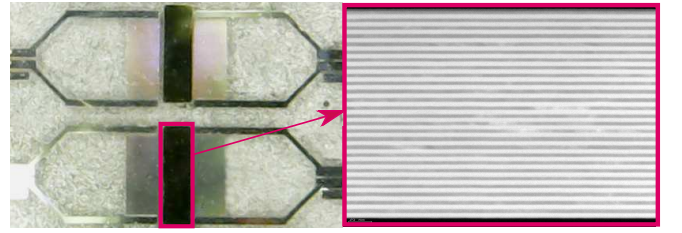}
    \caption{Photograph of the fabricated SAW delay line: the Al IDTs and the TbCo$_2$/FeCo multilayered thin film in-between (left). STEM picture of the multilayered TbCo$_2$/FeCo nanostructured thin film (overall thickness: 200 nm) deposited on silicon for Transmission Electron Microscopy visualization (right).}
    \label{fig:dispo}
\end{figure}
}

\subsection{Device characterization and Methods}
\begin{figure}[]
    \centering
    \includegraphics[width=\columnwidth]{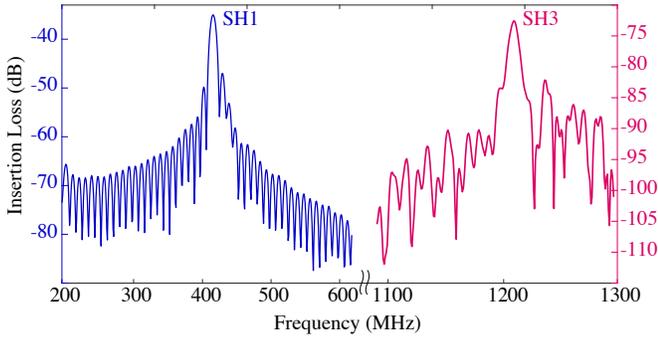}
    \caption{Frequency response of the realized SAW delay line on Quartz ST-90\textdegree{} X cut after deposition of the TbCo$_2$/FeCo thin film in the frequency range [200-1300] MHz showing the first (SH1) and third harmonics (SH3) of the shear-horizontal mode at zero magnetic field.}
    \label{fig:s21}
\end{figure}

The complex forward transmission $S_{21}$ of the delay line, was measured with a vector network analyzer (Agilent 8753ES) with an input power of P = 1 mW. The acoustic signal resulting from the SAW transmission can be isolated from spurious signals including electromagnetic crosstalk and multiple transit signals using Fourier transform and time gating, since the electromagnetic crosstalk is propagating at light speed and the surface acoustic wave five orders of magnitude slower. Therefore, these signals can be separated in time domain. The frequency response of the signal transmitted between the IDTs after thin film deposition is shown in FIG.~\ref{fig:s21}. The first (SH1) and third harmonic (SH3) of the pure shear-horizontal mode appear clearly on this electrical characteristic at 410~MHz and 1.2~GHz, respectively. In order to characterize the acoustic waveguide when the ferromagnetic thin film is biased with a magnetic field, the magnitude and phase value of the $S_{21}$ parameter are measured at the frequency showing the lowest insertion loss at zero field.

\begin{figure}
    \centering
    \includegraphics[width=\columnwidth]{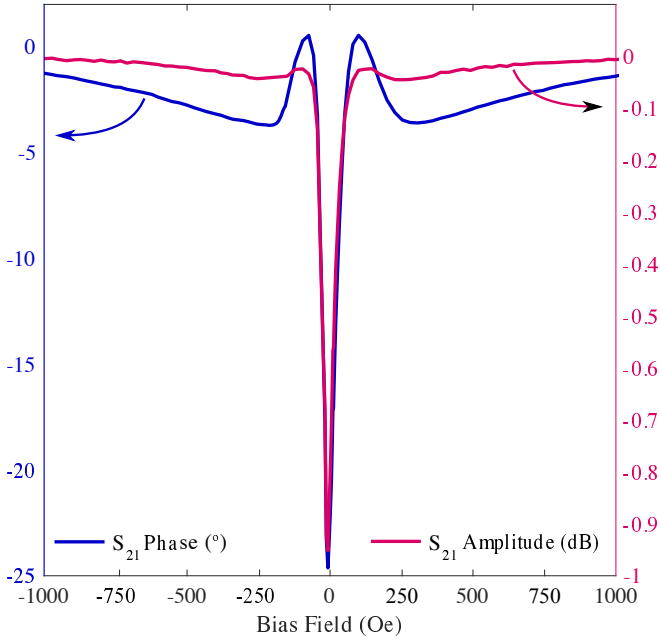}
    \caption{Complex $S_{21}$ response (phase in blue, magnitude in magenta) of the SAW delay line operated at 410 MHz as a function of the bias magnetic field applied along the hard axis of the multilayered TbCo$_2$/FeCo nanostructured thin film.}
    \label{fig:s21_fund}
\end{figure}

\hl{\subsection{Study of the SH1 mode at 410 MHz}
\label{disc}

\subsubsection{Bias magnetic field applied along the hard axis}

In a first set of measurements, the $S_{21}$ frequency response of the SH1 mode is measured as a function of the bias magnetic field, applied along the hard axis}. The normalized phase and amplitude of the insertion loss are depicted in FIG.~\ref{fig:s21_fund}. The amplitude shows a variation of nearly 1 dB for 100~Oe variation whereas the phase shift is around 25\textdegree{} for the same magnetic field range. Given the width of the magnetostrictive thin film (300~$\mu m$), the attenuation reaches 30~dB/cm. The $S_{21}$ amplitude variation can be explained by the modification of the acoustic impedance $Z=\rho c$ (with $\rho$, the thin film density and $c$ the wave propagation velocity), leading to a variation of the insertion loss with respect to the magnetic field.

\begin{figure}[]
    \centering
    \includegraphics[width=\columnwidth]{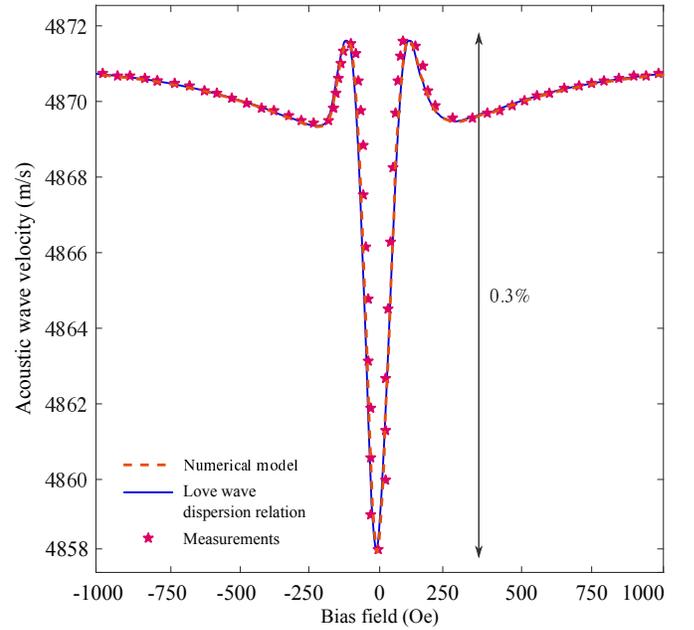}
    \caption{Comparison between measured SH$_1$ phase velocity, phase velocity shift obtained from Love wave dispersion relation and velocity shift obtained from FEM model when the multilayered TbCo$_2$/FeCo thin film is biased along its hard axis at 410 MHz.}
    \label{fig:vel}
\end{figure}

The $S_{21}$ phase shift is converted into a phase velocity shift through the relation
\begin{equation}
    \frac{\Delta v}{v} = -\frac{\Delta \Phi}{\Phi}
    \label{eq:phi2v}
\end{equation}
where $v$ is the phase velocity and $\Phi$ is the absolute phase.
In FIG.~\ref{fig:vel}, the resulting Love wave velocity behaviour of the SH1 mode (magenta dots) is compared to the Love wave velocity shift obtained through the dispersion relation (\ref{eq:lovewave}) by including $C_{66}$ and $C_{55}$ magnetic field dependency (blue). One can notice the perfect agreement between the velocity shift obtained with the established Love wave dispersion relation and the experimental measurements for the SH1 mode. The resulting velocity shift of the Love wave is 0.3\% at the fundamental frequency. 

\begin{figure}
    \centering
    \includegraphics[width=\columnwidth]{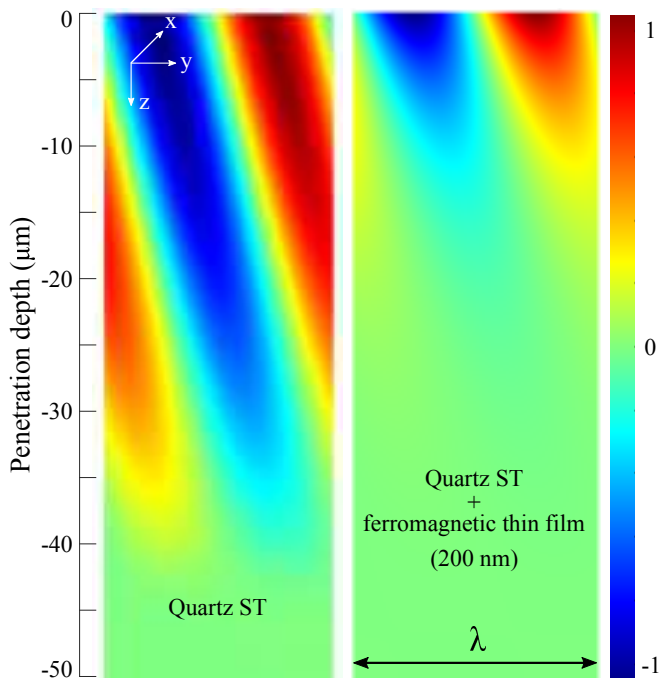}
    \caption{Normalized shear displacement field component computed with Comsol Multiphysics based on the equivalent piezomagnetic model when the ferromagnetic thin film is saturated.}
    \label{fig:comsol_disp_field}
\end{figure}

\hl{A numerical implementation of the equivalent piezomagnetic model was done in Comsol Multiphysics through Partial Differential Equation (PDE) module. The 3D unit cell, whose a 2D cut is represented in FIG.~\ref{fig:comsol_disp_field}, was constructed in Comsol and is composed of a block of Quartz ST-cut ($\lambda$ wide and $10\lambda$ high) coated with a block of the TbCo$_2$/FeCo ferromagnetic thin film (200 nm thick). Continuity condition was applied along the x and y directions and the bottom of the substrate was supposed to be fixed. The behaviour of the ferromagnetic thin film with respect to the bias magnetic field was obtained from the equivalent piezomagnetic model described in Section~\ref{eqpiezomag} and implemented as a new material. Therefore, the frequency of the shear eigenmode of the proposed structure was computed as a function of the bias magnetic field. The shear component of the displacement field is depicted in FIG.~\ref{fig:comsol_disp_field} for the SH1 mode. It is clearly shown that the coating with the ferromagnetic thin film (right) enhances the confinement of the acoustic wave in the upper layer compared to the bare Quartz ST-cut (left). The penetration depth is reduced by 1:4 at 410 MHz. The induced acoustic wave velocity shift with respect to the bias magnetic field is reported in FIG.~\ref{fig:vel} and compared to its evolution obtained with the analytical Love wave dispersion relation (\ref{eq:lovewave}) and with the experimental measurements. One may notice the good agreement between both methods proving the viability of the FEM tool to deal with the magnetoelastic coupling in more complex structures.}

\hl{\subsubsection{Influence of the bias magnetic field direction}}

\begin{figure}[]
    \centering
    \includegraphics[width=\columnwidth]{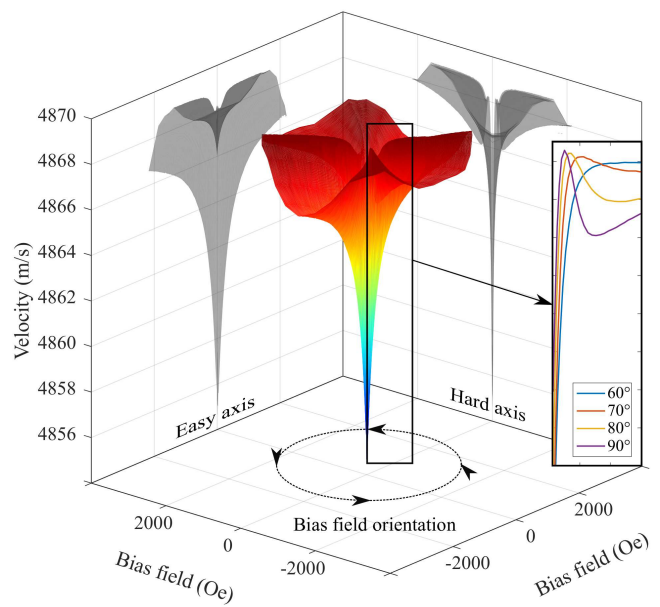}
    \caption{3D view of \hl{the experimental} phase velocity shift shape depending on bias magnetic field orientation. The inset shows overlaid cuts of the 3D figure for different orientations of the bias magnetic field. For the measurements carried along the hard and easy axis, the 3D figure is projected onto a plane (shaded figures) to highlight the velocity shift sensitivity with respect to the magnetic field.}
    \label{fig:3d}
\end{figure}

Magnetic field sensor response to differing bias magnetic field directions relative to the easy axis was investigated. The $S_{21}$ parameter of the SH1 mode was measured for bias magnetic field direction ranging from -90\textdegree{} to 90\textdegree{} with a step angle of 10\textdegree{} and a magnitude swinging from -2000~Oe 2000~Oe. The phase shift is converted into a velocity shift with the relation given in (\ref{eq:phi2v}) and the result is reported in FIG.~\ref{fig:3d} for the SH1 mode. As highlighted in the inset, as the bias magnetic field orientation gets closer to the hard axis, the velocity variation shows an increasingly recessed 'valley shaped' behavior. This is clearly the signature of the \textDelta E-effect \cite{robbins_simple_1988}, the velocity shift owing initial softening of elastic constants and then stiffening as applied field is increased. This valley shape is occurring around the anisotropy field of the multilayered ferromagnetic thin film close to 200~Oe. Indeed, in multilayered TbCo$_2$/FeCo ferromagnetic thin film exhibiting in-plane uniaxial anisotropy, a high susceptibility to the external driving field can be obtained in the vicinity of a field-induced Spin Reorientation Transition (SRT) \cite{Klimov2006} when the static external bias magnetic field is applied perpendicular to the easy axis with a magnitude equal to the anisotropy field $\textrm{H}_\textrm{A}$. Therefore, for a bias magnetic field close to that value, the elasticity of the magnetostrictive material becomes very soft due to magnetic in-plane rotation, hence the 'valley shape'.

\hl{\subsubsection{Considerations on Love wave guiding}}

\begin{figure}
    \centering
    \includegraphics[width=\columnwidth]{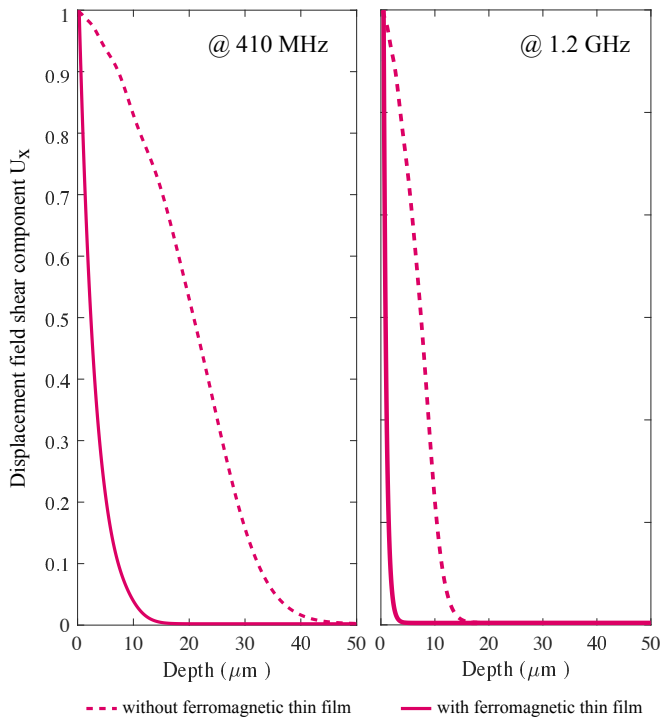}
    \caption{Comparison of the normalized displacement field component $U_x$ of the SH mode with or without the ferromagnetic layer coating (TbCo$_2$/FeCo) on quartz ST cut (if present, 200 nm) at two distinct frequencies, 410 MHz (left) and 1.2 GHz (right).}
    \label{fig:freq_comp}
\end{figure}

Typical Love wave devices reported in literature \cite{Kittmann2018, Liu2019} usually comes with a silicon dioxide guiding layer deposited on Quartz ST-cut to increase the wave confinement thanks to its lower shear velocity (close to 3764 m/s). In that case, the condition (\ref{eq:lwcond}) for Love wave to exist is fulfilled. In the proposed structure, the multilayered TbCo$_2$/FeCo thin film acts itself as a very effective guiding layer, its shear velocity being close to 1800 m/s (\ref{eq:vsh}), offering a higher contrast with the substrate shear velocity (close to 5000 m/s), compared to silicon dioxide. \hl{A numerical method that was developed in a previous work \cite{noauthor_legendre_nodate} to compute the dispersion curves and mode shapes of elastic waves in layered piezoelectric-piezomagnetic composites, is used to assess the effect of the ferromagnetic thin film coating (200 nm) on Quartz ST-cut. In FIG.~\ref{fig:freq_comp}, is reported the shear displacement component of the SH1 and SH3 modes corresponding to differing acoustic wave frequencies. The ferromagnetic layer clearly tends to confine the acoustic wave at the top of the structure. This is even more pronounced as the frequency increases, since the penetration depth is reduced to 1:4 at 410 MHz but 1:8 at 1.2 GHz. In comparison, the same effect is obtained with a 1.8 $\mu m$ thick silicon dioxide coating for the same Love acoustic wave velocity. Therefore, the wave energy is mainly confined in the upper layer, improving the surface acoustic wave interaction with the ferromagnetic thin film. Above all, it confirms that the ferromagnetic layer acts as a guiding layer for the acoustic waves.} 

\hl{These preliminary results carried on the SH1 mode show a very good agreement between the Love wave dispersion relation including the equivalent piezomagnetic model and the experimental measurements. As described in Section~\ref{eqpiezomag}, this model considers the magnetoelastic coupling occurring in ferromagnetic thin films and does not take into account any resonant coupling between spin wave and acoustic wave (known as ferromagnetic resonance). The ferromagnetic thin film shows a giant magnetoelastic coupling inducing large wave propagation velocity shift, but any evidence of the coupling between spin wave and acoustic wave was observed neither for the SH1 nor the SH3 mode. A possible reason may be linked to the hypothetical severe damping occurring in such ferromagnetic thin film, preventing the magnetization precession. Further investigations are required to shed light on the true coupling between dynamic strain and magnetization behaviour with respect to the magnetic field. As it stands, a pure magnetoelastic coupling model is perfectly able to explain what is observed experimentally in the proposed Love acoustic waveguide based on a TbCo$_2$/FeCo ferromagnetic thin film guiding layer.}

\hl{\subsection{Study of the SH3 mode at 1.2 GHz}
\subsubsection{Bias magnetic field applied along the hard axis}}

\begin{figure}[]
    \centering
    \includegraphics[width=\columnwidth]{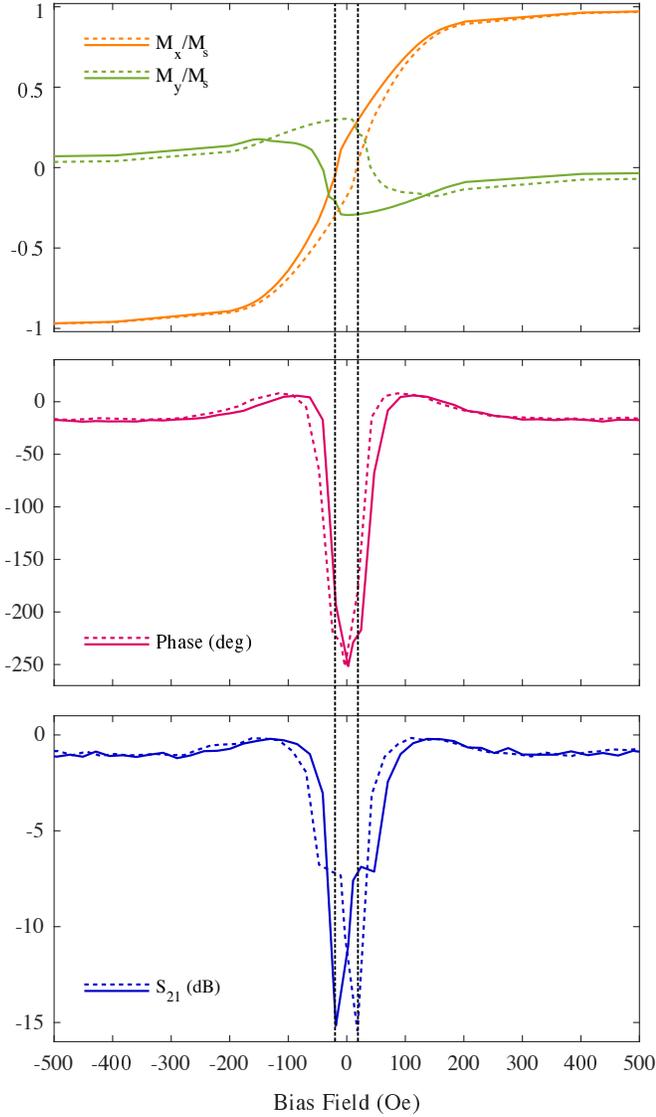}
    \caption{TbCo$_2$ /FeCo thin film magnetization characteristics along the hard axis (top). Normalized Insertion Loss ($S_{21}$) of the SAW delay line at 1.2 GHz (phase (middle) and magnitude (bottom)). In both figures, the dashed line represents measurements from left to right and solid lines from right to left starting from a saturated state.}
    \label{fig:mag_s21}
\end{figure}

\hl{The frequency response of the acoustic waveguide with respect to the magnetic field was also investigated for the SH3 mode at 1.2 GHz. The normalized phase and amplitude of the insertion loss are depicted in FIG.~\ref{fig:mag_s21} when the bias magnetic field is applied along the hard axis. The amplitude shows a variation of nearly 15~dB for 100~Oe variation whereas the phase shift is around 250\textdegree{} (middle) for the same magnetic field range. Given the width of the magnetostrictive thin film (300~$\mu m$), the attenuation easily reaches the value of 500~dB/cm (bottom). As a recall, the same measurements performed for the SH1 mode gave only 30~dB/cm of attenuation and 25\textdegree{} of phase shift. The $S_{21}$ variations with respect to the bias magnetic field are much more pronounced for the SH3 mode, due to a better confinement of the acoustic wave energy in the ferromagnetic thin film (as the frequency increases, the wavelength shortens) and thus, a more effective guiding leading to an increase in sensitivity.
One may also notice that the evolution of the $S_{21}$ parameter with respect to the bias magnetic field is closely related to the magnetization state of the ferromagnetic thin film. Vectorial measurements carried on the Vibrating Sample Magnetometer show the full magnetization behaviour when biased along the hard axis of the ferromagnetic thin film. As depicted in FIG.~\ref{fig:mag_s21} (top) where $M_x$ represents the magnetization along the bias direction (hard axis) and $M_y$ the simultaneous magnetization along the 90\textdegree{} direction, the magnetization is relaxing towards the easy axis from a saturated state along the hard axis in a 'Stoner Wohlfarth' fashion as it was discussed by Klimov \etal{} \cite{Klimov2006}}. As highlighted in FIG.~\ref{fig:mag_s21} (black dashed line), the hysteretic nature of the multilayered ferromagnetic thin film is reflected in the $S_{21}$ amplitude (bottom) and phase (middle). The minima are correlated to the 'jump' of the magnetization around zero field, both with increasing or decreasing magnetic field. This 'jump' occurring when the magnetization is relaxing from a saturated state along the hard axis corresponds to a brutal reorganization of the magnetization when crossing zero field as it was discussed by Tiercelin \etal{} \cite{Tiercelin2002}. Besides, it can be noticed that the fabricated SAW delay line is behaving like a bipolar magnetic field sensor. Indeed, the $S_{21}$ magnitude shows a crossing with opposite slope around zero field. Therefore, the SAW device is able to sense the direction of small magnetic field without any polarization.

\hl{\subsubsection{Acoustic mode conversion / Limitations of the analytical Love wave dispersion relation}

The Love wave velocity shift obtained experimentally or the SH3 mode is reported in FIG.~\ref{fig:vel_sh3} (magenta stars). The maximum velocity shift reaches 2.5\%, which corresponds, as the film thickness is only 0.8\% of the acoustic wavelength, to a variation of the elastic properties of the ferromagnetic thin film close to 100\%. 
One may further note that the 'valley shape' around the anisotropy field ($\textrm{H}_\textrm{A}$ = 200 Oe) is less recessed compared to the SH1 mode. The comparison between the experimental measurements and the velocity shift obtained with the Love wave dispersion relation reported in FIG.~\ref{fig:vel_sh3} (blue curve) shows that the analytical dispersion relation is not as accurate as it was for the SH1 mode to depict the evolution of the velocity with respect to the magnetic field close to the anisotropy field $H_A$ (200 Oe).

\begin{figure}[]
    \centering
    \includegraphics[width=\columnwidth]{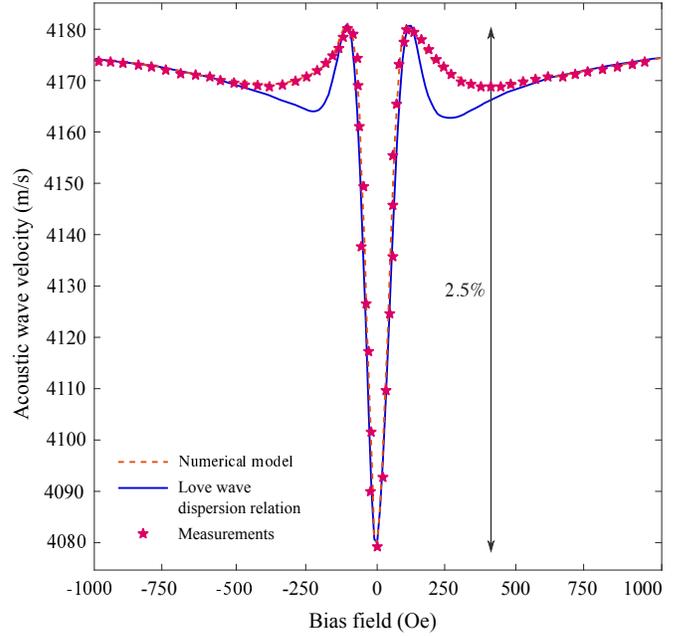}
    \caption{Comparison between measured SH3 phase velocity, velocity shift obtained from Love wave dispersion relation and velocity shift obtained from the numerical model when the multilayered TbCo$_2$/FeCo thin film is biased along its hard axis at 1.2 GHz.}
    \label{fig:vel_sh3}
\end{figure} 

Precisely, the analytical Love wave dispersion relation (\ref{eq:lovewave}) was obtained assuming that pure shear horizontal wave is propagating in the ferromagnetic thin film. This assumption is true in Quartz ST-cut (intrinsic to the material), but not completely when coated with the anisotropic ferromagnetic material. Indeed, as depicted in FIG.~\ref{fig:1200_comp}, the normalized displacement field components of the studied acoustic wave depend on the magnetic state of the ferromagnetic thin film and the frequency of operation. More precisely, it is shown that the ferromagnetic layer behaves as an isotropic layer when saturated and thus, the acoustic wave is pure shear (left). When the ferromagnetic thin film is not biased (zero field), its anisotropy gives birth to a three-component displacement field leading to a breaking of the acoustic wave polarization. As the frequency increases (middle and right), the two additional components become non-negligible and show more pronounced effect on the acoustic wave response with respect to the bias magnetic field. Precisely, for the SH3 mode, the contribution of the second component $U_y$ plays a role in the magnetoelastic coupling through the $C_{12}$ elastic stiffness constant, showing opposite behaviour compared to $C_{66}$ with respect to the magnetic field around the anisotropy field (200 Oe) as reported in Appendix~\ref{app:cij}. It may explain that the 'valley shape' is less recessed at 1.2 GHz, the $C_{12}$ contribution compensating the effect of the $C_{66}$. The anisotropic nature of the ferromagnetic thin film depending on the bias magnetic field, the acoustic wave polarization is linked to the bias magnetic field and therefore can lead to an acoustic mode conversion depending on the ferromagnetic thin film thickness and the acoustic wave frequency.

\begin{figure}
    \centering
    \includegraphics[width=\columnwidth]{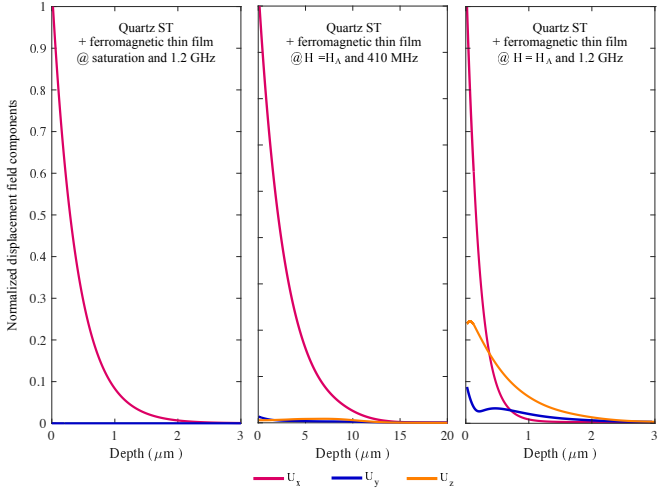}
    \caption{Comparison of the normalized displacement field components of the SH mode for [Quartz ST-cut + TbCo$_2$/FeCo multi-layer (200 nm)] at two distinct bias magnetic field, saturation field (left) and anisotropy field at 410 MHz (middle) and 1.2 GHz (right).}
    \label{fig:1200_comp}
\end{figure}

Therefore, the complete displacement field vector must be taken into consideration in order to depict the real behaviour of the Love acoustic wave with respect to the magnetic field. For that purpose, the implemented numerical model in Comsol Multiphysics can be used to compute the acoustic wave velocity shift as a function of the bias magnetic field. The results obtained with the numerical model are reported in FIG.~\ref{fig:vel_sh3}, showing a good agreement with the experimental data. One might further note that the velocity is lower as the frequency increases, proving that the acoustic wave is even more guided in the upper ferromagnetic layer, the velocity getting closer to the bulk shear velocity of the ferromagnetic thin film. Therefore, a frequency increase by a factor of three leads to an increase of the sensitivity by a factor of roughly ten. This behaviour is explained by the dispersive nature of the Love wave as the frequency increases. In order to assess the potential of the presented Love acoustic waveguide in terms of velocity shift with respect to the bias magnetic field (and thus, the sensitivity), the analytical dispersion relation is used. Using (\ref{eq:lovewave}), the maximum Love wave velocity shift with respect to the magnetic field is computed as a function of the ratio between the ferromagnetic film thickness $h$ and the acoustic wavelength $\lambda$. The maximum Love wave velocity shift is assessed using the relation (\ref{eq:dv}):

\begin{equation}
    \frac{\Delta v}{v_0}= \frac{v_{max} - v_{min}}{v_{max}}
    \label{eq:dv}
\end{equation}

\begin{figure}[H]
    \centering
    \includegraphics[width=\columnwidth]{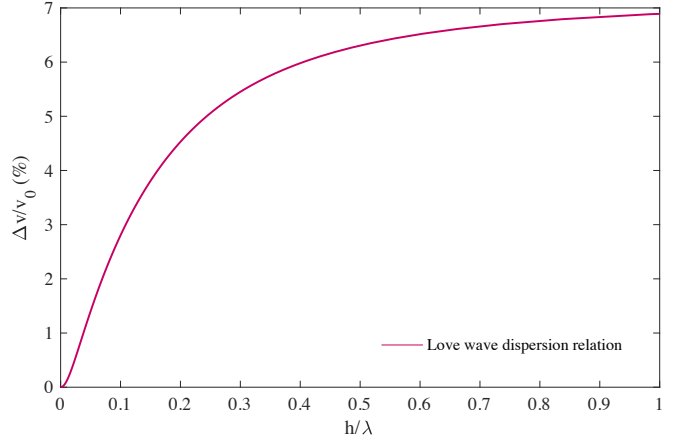}
    \caption{Love wave velocity shift with respect to the magnetic field as a function of the ratio between the ferromagnetic film thickness and the acoustic wavelength.}
    \label{fig:lw_sensitivity}
\end{figure}

The resulting sensitivity is depicted in FIG.~\ref{fig:lw_sensitivity}.
The maximum relative Love wave velocity shift is close to 7\% for a ratio $h/\lambda$ close to 1. The fabricated Love wave device reported in Section \ref{sec2}, shows only a ratio $h/\lambda$ close to 0.016, leaving room for improvement to design highly sensitive magnetic field sensor.
The analytical approach has the advantage of providing a complete understanding of the physical coupling, but the need of the global numerical model is required to take into account the true nature of the wave polarization as the frequency increases due to the anisotropic nature of the ferromagnetic thin film. The validity of the proposed equivalent piezomagnetic model proves to be satisfying even when working with complex wave polarization.}

\section*{Conclusion}
\hl{We presented here a thorough investigation, both theoretical and experimental, of giant magnetoelastic coupling in Love acoustic waveguides.
The existence of guided acoustic waves in a TbCo$_2$/FeCo ferromagnetic thin film on Quartz ST-90\textdegree X cut was demonstrated. The use of guided acoustic waves shows an exalted sensitivity by one order of magnitude compared to evanescent waves in conventional piezoelectric substrates. At a fundamental level, the proposed single-mode structure allows a complete understanding of the magnetoelastic coupling in such ferromagnetic materials and the development of an equivalent piezomagnetic model based on pure magnetoelastic coupling was used to assess the induced velocity shift due to \textDelta E/G-effect. Measurements performed on a SAW delay line coated with a uniaxial multilayered TbCo$_2$/FeCo nanostructured thin film were compared with very good agreement to the predictions of the model for the SH1 and SH3 modes. A maximum velocity shift of 2.5\% was reached for the SH3 mode. The attenuation induced by the ferromagnetic thin film in this configuration easily reached 500~dB/cm. The reported theoretical model and experimental results are of tremendous interest for the development of advanced devices for magnetic field sensing applications as well as investigating magnon-phonon interaction at a fundamental level. Finally, resonant coupling between magnon-phonon was not observed in our structure but may further improve the sensitivity.}\\

\begin{acknowledgments}

The authors would like to thank the support of the Agence Nationale de la Recherche through grant 2010 BLAN 923 0, RENATECH NETWORK and Centrale Lille through PhD fellowship.

\end{acknowledgments}

\appendix*
\section{\label{app:cij}}
\begin{figure*}
\includegraphics[width=\textwidth]{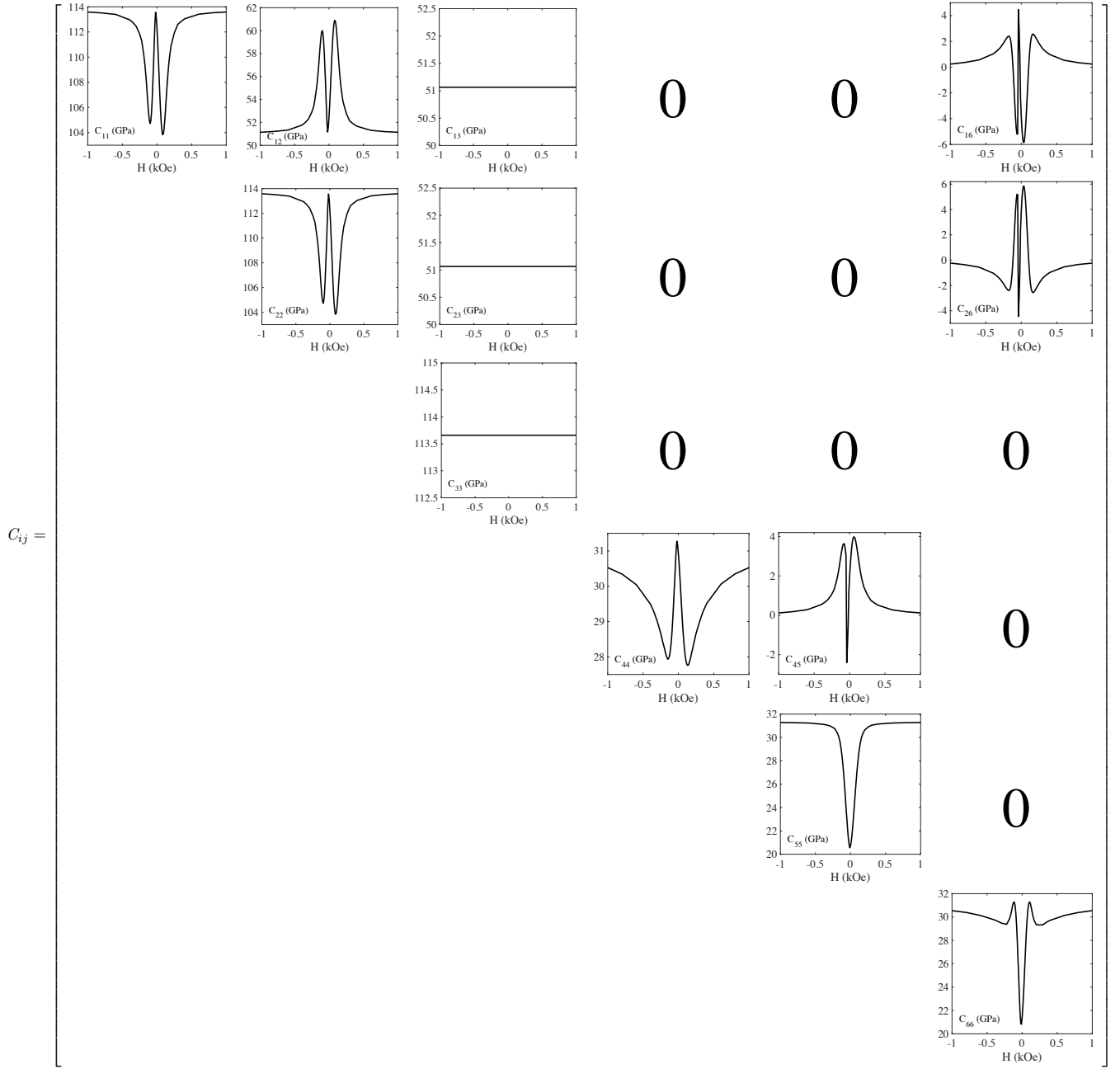}
\caption{Magnetic field dependency of the elastic stiffness constant tensor when a bias magnetic field is applied along the hard axis of a multilayered TbCo$_2$/FeCo thin film Quartz ST-X90\textdegree{} cut.}
\label{fig:cij}
\end{figure*}

The evolution of the elastic stiffness constants $C_{ijkl}$ of the ferromagnetic thin film with respect to the magnetic field is numerically computed using (\ref{eq:delta_cijkl}) and Table \ref{tab:deltac} and is reported in FIG.~\ref{fig:cij}. The magnetization state of the ferromagnetic thin film is taken into account through experimental measurements and implemented in the model. The reported results are obtained for a bias magnetic field applied along the hard axis of the ferromagnetic thin film.

\clearpage
\bibliographystyle{ieeetr} 
\bibliography{References}

\end{document}